\documentclass[prb,aps,superscriptaddress,endfloats, showpacs]{revtex4}

\begin {document}
\Large {
\title {Relativistic Quantization of Cooper Pairs and \\ Distributed Electrons in Rotating Superconductors}

\author {I.E. Bulyzhenkov} 
 \affiliation 
 {P.N. Lebedev Physical Institute, Leninsky pr. 53, Moscow, 119991, Russia}


\bigskip
\begin {abstract}
{\large Relativistic time synchronization along  closed integral lines maintains magnetic flux quantization independently from gravitation. 
All Fermi-volume electrons form time-averaged electromagnetic fields within 
rotating conductors, while Fermi-surface superelectrons enable flux quantization in SQUID experiments. Inertia is not related to instantaneous self-coherent states  of the distributed electric charge and, therefore, the Cooper pair mass can not be measured in principle from magnetic flux quantization.     }
\end {abstract}
 \bigskip

\maketitle

\bigskip

\bigskip

The known experimental efforts \cite{Jtate, Tat, Tate} to determine `the observable Cooper pair mass'  by SQUID has not been satisfactorily commented by  theorists. The well measured family of magnetic flux outputs versus rotation rates of the niobium ring at 6 K is  still unexplained in  conventional terms. This provisional uncertainty cannot  terminate speculations about special relativistic options for accelerated superconductors and their `prospective' inertial tools like quantum accelerometers. Here we start with  Feynman's path integrals in curved 4D  geometry for quantized potential states of a nonlocal charged particle. Then we apply the instantaneous relativistic quantization to nonlocal distributions of the continuous electron and the continuous Cooper boson in  a superconductor with the rotationally induced London moment ${\bf B }\propto  $  {\boldmath$\omega $}. The physical origin of this inertial field is related to time-averaged motion of ideal Fermi-liquid in accelerated electroneutral conductors or rotating Faraday disks. Below the superconducting transition temperature this ideal fermion fluid gains a small fraction of superfluid pairs, which does not change bulk induced fields but enables magnetic flux quantization in macroscopic hollow cylinders and rings. Such time-independent quantization due to a steady self-coherent state of the continuously distributed Cooper boson does provide precision SQUID measurements of minimal flux changes for the exact determination of the superfluid particle charge,  but not for determinations in question of the Cooper pair mass or inertia.

Potential motion of two paired superelectrons can be described
by one scalar complex field    $\varphi$ = $|\varphi| exp (i\chi/\hbar c)$ for this continuously distributed formation of charged matter. All Cooper pairs have the same charge $Q$ and identical relativistic energies $Mc^2$ due to pairing on the same Fermi surface $\epsilon_{_F} $.  High Fermi surface speeds, $v_{_F} \approx 10^6 m/s$, of superelectrons may provide high relativistic corrections, $(M-2m_o)/2m_o = \epsilon_{_F}/m_o c^2 \approx 10^{-4}$, to the expected inertia  $M = 2m(\epsilon_{_F})  \equiv 2m_o/{\sqrt {1- v^2_{_F}c^{-2}}}  \equiv 2m_o + c^{-2}\epsilon_{_F}  $ of the Cooper pair `at rest' (or at low speeds of rotating superconductors). The reported `observable mass' $M^* = 2m_o \times 1.000084(21)$
from the flux data interpretation \cite{Tat, Tate} has  been  agreed neither with the pure mechanical energy prediction for two superelectrons, $ 2m(\epsilon_{_F})/2m_o = 1.000180$ due to the niobium  Fermi energy  $\epsilon_{_F} = 0.000180 m_oc^2$, nor with the electrochemical energy calculations, $2m'/2m_o = 0.999992$ due to the negative electrostatic potential of Fermi surface electrons within the niobium lattice. 

How to understand what, in fact, was found in these well-performed SQUID measurements? At first, we discuss inertial contributions into induced bulk fields of relativistic ideal fluid of nonsuperconducting electrons, which occupy different energy levels under the Fermi surface. Then we prove quantitatively that electrical (negative or positive)  potential energy is not relevant to flux quantization of massive charges (otherwise one could expect Thomson's contribution of electricity into inertia of a charged corpuscle). After this we propose the relativistic generalization of the Bohr-Sommerfeld quantization rule without any contributions from gravitational potentials.  

In order to describe the relativistic potential motion of superfluid particles in curved space-time,  for example \cite{Bul},  one may relate their  canonical four-momentum density, $n(x) P_\mu (x) = -  n(x)\nabla_\mu \chi /c \equiv - n\partial_\mu \chi /c $, to the phase function gradient,  where $\nabla_\mu $ is the covariant derivative and
$\mu \rightarrow \{ 0,1,2,3\}$. The basic covariant equality $\nabla_\mu P_\nu -
\nabla_\nu P_\mu \equiv \partial_\mu P_\nu -
\partial_\nu P_\mu \equiv 0$ for the canonical four-momentum $P_\mu$ in metric fields with symmetrical Christoffel
coefficients seems formally independent from the material density distribution $n = n(x)$. However, these continuous density must be finite in all space points where one defines the single-valued phase $ \chi $ (otherwise the phase-matter relation reads $0 = 0$). Therefore, the phase equality 
 $\partial_\mu \partial _\nu  \chi \equiv  \partial _\nu \partial _\mu \chi  $ can be applied to continuous superfluid matter only in material points, where $n(x)\neq 0$. When one applies this  `trivial' equality to a local phase $\chi_e (x)$ of one self-coherent electron (in the absence of energy exchanges with other particles), then this electron should be also considered as a continuous distribution with a finite material density, $n_e(x) \neq 0$, instantaneously in all space points. We assign a finite canonical energy-momentum density to all space points of the continuous elementary particle (without spin) in gravitational and electromagnetic fields,   
\begin {equation}
 - {n_e\over c}\partial_\mu \chi_e  = n_e P_\mu \equiv n_e g_{\mu \nu }{\left (m_o  c
{dx^\nu\over  ds} + {{e }\over c}A^\nu \right )}  \equiv \{ n_e P_o;  n_eP_i\}$$ $$
  \equiv \left \{ \left [ {\frac {n_e m_o c {\sqrt {g_{oo}}}} {\sqrt {1 - v^2c^{-2}}}} + {  g_{o\nu}n_e eA^\nu \over c}
\right ];
 \left [ {\frac  { - n_em_o( \gamma_{ij} v^j +  {\sqrt {g_{oo}}} g_i c)} {\sqrt {1 - v^2c^{-2}} } } + { g_{i\nu}n_eeA^\nu\over c} \right ] \right \}. 
\end {equation}
Hereinafter we use the relativistic three-velocities 
$v^i \equiv cdx^i (g_{oo})^{- 1/2}(dx^o - g_idx^i)^{-1} \equiv dx^i /d\tau$ ($v^2 \equiv \gamma_{ij}v^iv^j$,
 $\gamma_{ij} \equiv g_ig_jg_{oo}- g_{ij}$, $g_i \equiv - g_{oi}/g_{oo} $, $i \rightarrow \{1,2,3\}$) of the continuous material density $n_e$. In other words, Bohm's interpretation of distributed self-coherent electron's states is more close to our relativistic approach, than Born's probability dice for the point electron.

 The basic covariant equalities, $\partial_\mu \partial _\nu  \chi_e \equiv  \partial _\nu \partial _\mu \chi_e  $, for the electron's phase  accompany the potential motion  (between collisions) of  electron's distributed densities $ m_o n_e$ and $e n_e$ in external  electric, ${ E_i}  \equiv - { \partial_i }(g_{o\mu}A^\mu) - { \partial_o }(g_{i\mu}A^\mu) $, and magnetic, ${ B^i} \equiv - e^{ijk}[\partial_j (g_{k\mu} A^\mu) -  \partial_k (g_{j\mu} A^\mu)]/2 {\sqrt {|\gamma_{lm}|}} $, fields. By taking these definitions for components of relativistic fields, one can rewrite the basic phase equalities for continuous mass and charge densities,  $m_o n_e (x)$ and $e n_e (x) $, of the distributed self-coherent electron,   
\begin{eqnarray}
E_i  = 
 { \partial \over \partial x^i }  \left [{\frac {m_on_e c^2 {\sqrt {g_{oo}}} } {en_e\sqrt {1 - v^2c^{-2}}} }
 \right ]  + 
 { \partial\over \partial x^o} \left [{\frac {m_on_e c  (\gamma_{ij}v^j + {\sqrt {g_{oo}}} g_i c  )  } {en_e\sqrt {1 - v^2c^{-2}}}    }  \right ],
  \\ \nonumber
{\bf B}  = -  curl \left [{{m_o n_ec  \left ( {\bf v} + c{\sqrt {g_{oo}}}{\bf g} \right ) } \over {en_e{\sqrt {1 - v^2c^{-2}}}}}   \right ].   
\end{eqnarray}

All mobile charges with inertia can induce within an accelerated electroneutral conductor local electromagnetic fields, $E_i $ and $B_i$, for a probe charge.  For example, a constant angular velocity {\boldmath $\omega$}  of a rotating niobium ring results in steady bulk fields, $E_i = const$ and $B_i = const$, which correspond to the constant time-averaged  velocity, $<{\bf v}>_t $
= {\boldmath $\omega$}$\times$${\bf r }$, 
$ \omega r << {\sqrt {<v^2>_t} } << c $, and constant time-averaged  inertia   $<m(\epsilon) >_t = m_o<(1 - v^2c^{-2})^{-1/2}>_t \approx m_o[1 + ( <v^2>_t/2c^2)]$ in a laboratory system, where $g_i = 0$. All Fermi-liquid electrons may undergo  energy exchanges and phase memory breaking after mutual replacements inside the Fermi energy volume. Then, a dissipation-free system of fermionic loops over a hollow cylinder will not exhibit quantized time-averaged flux,  because $<\chi_e>_t = 0$ for every fermion and, consequently, for the ideal fluid of normal electrons. Their mutual replacements can be associated with quantum fluctuations at zero temperature and can be facilitated by Fermi volume holes at finite temperatures. One can approximate the time-averaging of electron's kinetic energy, $m_o<v^2>_t/2 \approx \epsilon_{_F}/2$, by its averaging over all Fermi energy levels due to  short-time fluctuation replacements within the Fermi volume, where  $0 \leq m_ov^2/2 \leq m_ov_{_F}^2/2 = \epsilon_{_F}$. This universal averaging for each and all Fermi volume electrons corresponds to their uniform rotation in inertial electric and magnetic,
 \begin {equation}          
{\bf B}  = - \frac {2 m_o c} {e} [1 + (\epsilon_{_F}/2m_o c^2) ] {\mbox{\boldmath$\omega$}},
\end {equation}
fields, despite charged fermions always have slightly  different relativistic inertia-energy. It is this difference which costs phase memory and flux quantization losses to free Fermi liquid even in ideal conductors with zero resistances.  
Bose condensation of some Fermi-surface electrons into Cooper pairs cannot change the net inertial field (3), but can provide steady 
superfluid loops with $<\chi_s(x)>_t \neq 0 $ over a hollow cylinder. These overlapping loops maintains magnetic flux quantization for the elementary  charge Q of each boson with identical energies.

Now we generalize the Bohr-Sommerfeld quantization rule on the canonical four-momentum density (1) of a distributed self-coherent particle with  continuous mass, $mn(x)$, and charge, $qn(x)$, densities by taking into account the General Relativity time synchronization, 
$d\tau \equiv {\sqrt {g_{oo}}}(dx^o - g_idx^i) = 0$ or $dx^o = g_idx^i$, for neighboring space points in Feynman's path-integrals, 
\begin{equation}
 {1\over c} \oint_{d\tau = o} dx^\mu{{\nabla_\mu \chi_s } } =  \oint_{d\tau = o} \left[- P_idx^i - P_o dx^o\right] 
 \equiv$$ $$\oint_{d\tau = o}\left[  {\frac  {m({\gamma_{ij}v^j}+{c\sqrt {  g_{oo}}}g_i)}{\sqrt {1 - v^2 c^{-2} }} } - { q(g_{io}A^o+g_{ij}A^j)\over c}-
 \left(   {\frac {mc\sqrt {g_{oo}}}  {\sqrt {1 - v^2 c^{-2} }} } + { q(g_{oo}A^o+g_{oj}A^j )\over  c}\right )g_i\right ]dx^i
$$ $$\equiv  \oint_{d\tau = o}   \left [{qA^j\over c}  + {\frac  {m{v^j }}{\sqrt {1 - v^2 c^{-2} }} }   \right ] \gamma_{ij} dx^i = \pm 2\pi N\hbar .
\end{equation}
Coulomb and Newton potentials do not contribute to this quantization for instantaneous relativistic distributions of particle's charge and mass densities in electromagnetic and gravitational fields. 
However, this rule was formally derived for gravity-dependent 3D space with the inhomogeneous metric tensor
 $\gamma_{ij}(x)\equiv g_ig_j g_{oo} - g_{ij}$. Path-dependent phase shifts contradict to single valued coherent states, Bohr-Sommerfeld quantization, and the path-integrals formalism for Feynman's quantum mechanics. Where is a possible solution for this provisional conflict between relativistic and quantum theories?  

In our view, Einstein's General Relativity can rigorously replace \cite{Buly} the operator mass  density $m\delta ({\bf x}- {\bf X}_k)$ of the postulated point particle with an analytical density  solution for a radial electron $ m_o n_e({\bf x}, t) = m_o r_e /4\pi ({\bf x}- {\bf X}_e)^2 (r_e + |{\bf x}-{\bf X}_e|)^2$, where $r_e = Gm_o/c^2$ in such a probability-free approximation of continuous elementary matter around its center of symmetry ${\bf X}_e = {\bf X}_e(t)$.  A similar analytical solution $ e n_e({\bf x}, t)$ to Maxwell's equations can be also employed for the electron's continuous charge. Moreover, the warped 4D interval for distributed radial densities  $mn({\bf x}, t)$ of particles or sources gains six universal  symmetries, $\gamma_{ij} (x) = \delta_{ij}$, for the 3D sub-interval in Einstein's nonlocal gravitation \cite{Buly}. Such universal spatial flatness in covariant equations reinforces the surface independent magnetic flux and its gravity/inertia independent quantization,   
         \begin{equation}
  \oint_{d\tau = o} {\delta_{ij}} q  A^j dx^i    \equiv  q \int_{d\tau = o} {\bf B} d{\bf s} = \pm 2\pi N \hbar c + m c^2 \oint_{d\tau = o}  idl,       \end{equation}
for instantaneously closed contours with
  $d\tau = 0$ and virtual displacements $idl$ of the inertial energy, $mc^2 \delta_{ij}v^j dx^i/(1 - v^2c^{-2})^{1/2} \equiv mc^2 dl^2/d{\sqrt {d\tau^2 - dl^2}} \equiv mc^2(-i dl) $, at any fixed moment of time.

The canonical four-momentum density (1) can be applied to the self-coherent Cooper pair with continuous charge $Qn_s(x)$ and mass $Mn_s(x)$ densities in potential fields. The pair is not involved in energy exchanges and obeys the magnetic flux quantization (4)-(5) during long laboratory measurements.  Time-varying changes of the steady field (3) in (5) are associated with frequency shifts $\Delta \nu = \Delta \omega /2 \pi$.  
By comparing neighboring steady states in (5), one should drop impossible (imaginary) mass replacements at zero time rate, $d\tau = 0$,  and infer that the Cooper pair charge Q was determined in the SQUID experiments
 with the record accuracy, 
 \begin {equation}
\left |\frac {Q}{2e} \right |  =   \frac {{ \hbar }/{ 4 m_o S  \Delta \nu}}{[1 + (\epsilon_{_F}/2 m_oc^{2})]} =   \frac {1.000084(21)}{1.000090} = 0.999994(21),    
    \end {equation}
  Here $S$ and $\Delta \nu$ were measured, while $\epsilon_{_F}$ was evaluated for niobium from independent sources \cite {Jtate, Tat, Tate}.
 The available data  quantitatively confirm that inertially induced fields within accelerated superconductors originate from the ideal averaged motion of all Fermi-volume electrons. According to (6), the Cooper pair charge can be formed as by two electrons, $Q=2e < 0$, as well as by two holes, $Q= -2e > 0$. 
  
  There is no mystery in the failed attempts to measure the relativistic Cooper  pair mass. Relativistic quantization (4)-(5) is completely free from gravitational/inertial fields. Inertia of continuous charges   is  irrelevant to their instantaneous self-coherent distributions in external electromagnetic fields. Were 3D space warped,   $\gamma_{ij}(x) \neq\delta_{ij}$, then gravity/acceleration control of quantized flux (5) in superconducting rings might be expected for a moment. But gravitational analogs of the Aharonov-Bohm effect were never reported in SQUID measurements that is in full agreement with the flatspace reading \cite{Buly} of Einstein's relativistic physics through nonlocal radial particles.

}
\end {document}